\documentclass[a4paper,twocolumn,english,american,prl,superscriptaddress,longbibliography,fixfloat,notitlepage]{revtex4-1}
\usepackage[latin9]{inputenc}
\usepackage{xcolor}
\usepackage{pdfcolmk}
\usepackage{amsmath}
\usepackage{amssymb}
\usepackage{graphicx}
\PassOptionsToPackage{normalem}{ulem}
\usepackage{ulem}

\makeatletter

\pdfpageheight\paperheight
\pdfpagewidth\paperwidth

\providecolor{lyxadded}{rgb}{0,0,1}
\providecolor{lyxdeleted}{rgb}{1,0,0}

\DeclareRobustCommand{\lyxsout}[1]{\ifx\\#1\else\sout{#1}\fi}

\usepackage{xcolor}
\definecolor{darkblue}{rgb}{0.1,0.2,0.6} 
\definecolor{lightblue}{rgb}{0.1,0.1,1.0}
\definecolor{darkred}{rgb}{0.8,0.1,0.2}
\usepackage[colorlinks,citecolor=lightblue,linkcolor=darkblue,urlcolor=lightblue] {hyperref}
\usepackage[colorlinks]{hyperref}

\makeatother

\usepackage{babel}
\begin{document}
\title{Noise-induced transport in the Aubry-André-Harper model}
\author{Devendra Singh Bhakuni}
\affiliation{Department of Physics, Ben-Gurion University of the Negev, Beer-Sheva
84105, Israel}
\author{Talía L. M. Lezama}
\affiliation{Department of Physics, Yeshiva University, New York, New York 10016,
USA}
\author{Yevgeny Bar~Lev}
\affiliation{Department of Physics, Ben-Gurion University of the Negev, Beer-Sheva
84105, Israel}
\email{ybarlev@bgu.ac.il}

\begin{abstract}
We study quantum transport in a quasiperiodic Aubry-André-Harper (AAH)
model induced by the coupling of the system to a Markovian heat bath.
We find that coupling the heat bath locally does not affect transport
in the delocalized and critical phases, while it induces logarithmic
transport in the localized phase. Increasing the number of coupled
sites at the central region introduces a transient diffusive regime,
which crosses over to logarithmic transport in the localized phase
and in the delocalized regime to ballistic transport. On the other
hand, when the heat bath is coupled to equally spaced sites of the
system, we observe a crossover from ballistic and logarithmic transport
to diffusion in the delocalized and localized regimes, respectively.
We propose a classical master equation, which captures our numerical
observations for both coupling configurations on a qualitative level
and for some parameters, even on a quantitative level. Using the classical
picture, we show that the crossover to diffusion occurs at a time
that increases exponentially with the spacing between the coupled
sites, and the resulting diffusion constant decreases exponentially
with the spacing.
\end{abstract}
\maketitle

\section{Introduction}

Non-equilibrium dynamics of quantum systems has been a topic of great
interest in condensed matter and statistical physics, particularly
concerning the study of quantum transport \citep{rigol2008thermalization,polkovnikov2011nonequilibrium,Nandkishore:rev,bar_lev_absence_2015,agarwal_anomalous_2015,luitz_ergodic_2017,agarwal_rare-region_2017,agarwal_localization_2017}.
While transport in generic many-body systems is typically diffusive,
diffusion of a quantum particle can be suppressed by a random disordered
potential due to localization of the single-particle states. This
phenomenon, dubbed Anderson localization \citep{anderson_absence_1958},
occurs for any non-zero random disorder in one and two dimensions.
At higher dimensions, \emph{all} states are localized only for sufficiently
high disorder. For disorder values smaller than a critical value,
only eigenstates below a specific energy, called the mobility edge,
are localized \citep{abrahams_scaling_1979,evers_anderson_2008,kramer_localization_1993}.

Localization also occurs for systems with non-random, quasi-periodic
potentials. Such potentials can exhibit localization-delocalization
transitions and mobility edges, even in one dimension \citep{aubry1980analyticity,harper_1955,biddle2011localization,ganeshan_nearest_2015,deng_many-body_2017,li2017mobility,wang2020one,duthie2021self},
and are experimentally feasible \citep{luschen_observation_2017,luschen_signatures_2017,luschen_single-particle_2018,an2021interactions}.
In the quasi-periodic Aubry-André-Harper (AAH) potential, for example,
all states are localized above a critical potential strength and delocalized
otherwise, whereas the transition point features fractal states \citep{aubry1980analyticity,harper_1955}.
Transport is absent in the localized phase, ballistic in the delocalized
phase, and anomalous at the critical point due to the fractal structure
of the eigenstates \citep{varma_fractality_2017,purkayastha_anomalous_2018,ng_wavepacket_2007,ketzmerick_what_1997}.

\begin{figure}
\includegraphics[width=0.9\columnwidth]{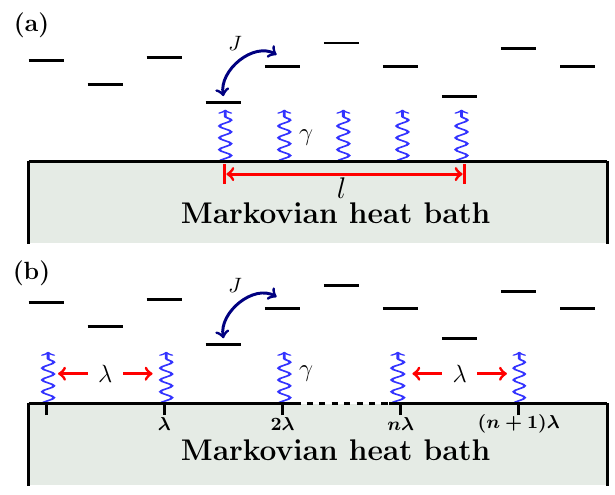}

\caption{Schematic of the coupling of the heat bath to the system. (a) central
region of $l$ sites is coupled (b) sites at $\lambda$ distance apart
are coupled.}
\label{fig:sch}
\end{figure}
In recent years a number of theoretical works suggested that single-particle
localization with or without quenched disorder is stable to the addition
of sufficiently weak inter-particle interactions, giving rise to novel
dynamical phase transitions away from thermal equilibrium \citep{basko_metalinsulator_2006,Gorny:2005,iyer_many-body_2013,imbrie_many-body_2016,Nandkishore:rev,Alet:rev,Abanin:rev}.
This phenomenon, dubbed many-body localization (MBL), manifested in
the absence of all transport for a sufficiently large disorder, though
logarithmically slow transport in the localized phase was reported
recently and is still under debate \citep{kiefer2020evidence,luitz2020absence,maximilian2022particle}.
For weak disorder, there has been numerical evidence of anomalous
diffusion that is attributed to the existence of rare insulating regions
\citep{bar_lev_absence_2015,agarwal_anomalous_2015,luitz_ergodic_2017,agarwal_rare-region_2017}.
The anomalous transport was also observed in quasiperiodic systems
even though such systems do not exhibit rare regions due to the deterministic
nature of the potential \citep{bar_lev_transport_2017,xu_butterfly_2019,yoo_nonequilibrium_2020,znidaric_comment_2021}.

The stability of the many-body localized phase was recently challenged
in a number of analytical and numerical works \citep{suntajs_quantum_2020,Kiefer:2021,Sels:2021,sierant_challenges_2022,Sels2022bath},
which argue that the MBL phase is unstable in the thermodynamic limit
and at long times. One proposed mechanism for destabilizing a many-body
localized phase in disordered systems is through avalanche propagation
\citep{Roeck:2016,deroeck2017stability,luitz_bath:2017,Thiery:2018,Crowley:2020,Morningstar:2022,Leonard:2023},
where small thermalizing regions known as thermal inclusions act as
a bath, which grow and take over the localized regions, leading to
delocalization of the system. The impact of these rare regions, also
called ergodic bubbles, have also been investigated in cold-atom experiments
\citep{luschen_observation_2017,Leonard:2023}.

Various phenomenological models of ergodic bubbles were proposed:
such as random matrices \citep{schiro2020toy} or coupling to a bath
\citep{nandkishore_bath:2014,luitz_bath:2017,deroeck2017stability,taylor_subdiffusion_2021,turkeshi_destruction_2022,lezama_mergold_love_logarithmic_2022,Brighi:2022,Zhou:2022}.
Coupling a Markovian bath to \emph{all} lattice sites of an Anderson
insulator is known to destroy localization and results in diffusive
transport~\citep{znidaric_2010,znidaric_2013,Moix_2013,medvedyeva_2016,znidaric_2017},
while for correlated noise, a transient sub-diffusive regime can be
seen \citep{gopalakrishnan_noise-induced_2017}. On the other hand,
it was demonstrated that local coupling to a Markovian bath induces
logarithmic transport in an Anderson insulator \citep{lezama_mergold_love_logarithmic_2022}.

The stationary current in a system with a finite density of sites
coupled to a heat bath, was studied inRefs.~\citep{taylor_subdiffusion_2021,turkeshi_destruction_2022}.
These works established a transition from a super-diffusive to diffusive
stationary current as a function of the density of the coupled sites.
However, Refs.~\citep{taylor_subdiffusion_2021,turkeshi_destruction_2022},
do not provide insight on the temporal dependence of spreading excitations.
Moreover, there are known cases where the behavior of the stationary
current is different from temporal spreading of density excitations~
\citep{varma_fractality_2017,purkayastha_anomalous_2018}. It is therefore
an open question if the spreading of density excitations exhibit one
or several regimes of transport, and how these regimes depends on
the coupling to the heat bath.

In this work, we answer these questions by studying the spreading
of density excitations at infinite temperature for different couplings
of the system to a Markovian bath.

The paper is organized as follows. We describe the model Hamiltonian
and the methods in Section~\ref{sec:Model-Hamiltonian-and}. In Section~\ref{sec:Results}
we present our results for different couplings of the model to a Markovian
heat bath. Finally, we discuss our findings in the Section~\ref{sec:Conclusions-and-Discussions}.

\section{Model and methods \label{sec:Model-Hamiltonian-and}}

We consider a system of $N$ spinless fermions in a chain of length
$L$, which is described by the Hamiltonian,
\begin{align}
\hat{H}= & -J\sum_{m=1}^{L-1}\left(\hat{a}_{m+1}^{\dagger}\hat{a}_{m}+\hat{a}_{m}^{\dagger}\hat{a}_{m+1}\right)\label{eq:hamiltonian}\\
+ & W\sum_{m=1}^{L}\cos\left(2\pi\beta m+\phi\right)\hat{n}_{m},\nonumber 
\end{align}
where $\hat{n}_{m}=\hat{a}_{m}^{\dagger}\hat{a}_{m}$, and $\hat{a}_{m},\hat{a}_{m}^{\dagger}$
are annihilation and creation operators of a fermion on site $m$,
$J$ is the hopping strength, $W$ is strength of the potential and
$\beta=\left(\sqrt{5}-1\right)/2$ is the Golden mean. The phases
are taken uniformly from $\phi\in\left[-\pi,\pi\right].$ In contrast
to the Anderson model, where all the single-particle eigenstates are
localized for any non-zero value of $W$, the AAH model exhibits a
delocalization-localization transition. For $W<2J,$ \emph{all} the
single-particle eigenstates are extended and transport is ballistic
\citep{purkayastha_anomalous_2018}, while for $W>2J$ \emph{all}
the states are localized. At the critical point, $W=2J$, the states
are multi-fractal and transport is diffusive if characterized by the
mean squared displacement \citep{purkayastha_anomalous_2018}, while
a characterization based on the stationary current suggests a sub-diffusive
transport~\citep{varma_fractality_2017,purkayastha_anomalous_2018}.

We couple the system to a Markovian heat bath which does not affect
the number of fermions in the system. Specifically, we assume that
the density matrix of the system evolves via the Lindblad master equation
\citep{lindblad1976generators},
\begin{align}
\frac{\partial\hat{\rho}\left(t\right)}{\partial t}= & -i\left[\hat{H},\hat{\rho}\left(t\right)\right]+\sum_{i}\left(\hat{L}_{i}\hat{\rho}\left(t\right)\hat{L}_{i}^{\dagger}-\frac{1}{2}\left\{ \hat{L}_{i}^{\dagger}\hat{L}_{i},\hat{\rho}\left(t\right)\right\} \right),\label{eq:lindblad}
\end{align}
where $\left\{ .\right\} $ is the anti-commutator, the first term
represents the unitary evolution and the second term gives the non-unitary
dynamics. The operators $\hat{L}_{i}$ are Lindblad jump operators,
which we take to be
\begin{equation}
\hat{L}_{i}=\sqrt{\gamma_{i}}\hat{n}_{i},\label{eq:lindblad-jump-ops}
\end{equation}
where $\gamma_{i}$ represents the strength of the dissipation on
site $i$. The dimensions of the density matrix are $\mathcal{N}\times\mathcal{N}$,
where $\mathcal{N}$ is the Hilbert space dimension. This unfavorable
scaling with the system size makes the numerical solution of the Lindblad
equation computationally expensive. A more efficient approach is to
unravel the evolution into a \emph{unitary} evolution of wavefunctions
in the presence of white noise and then to average over the realizations
of the noise to obtain the quantities of interest~\citep{Brun_2000,wiseman_2001,gardiner2004quantum,chenu_quantum_2017}.

For unitary unraveling, the time evolution operator is given by,

\begin{align}
\hat{U}(t+dt,t)= & e^{-i\hat{H}dt-i\sqrt{\gamma\,dt}\sum_{i}\eta_{i}\left(t\right)\hat{n}_{i}},
\end{align}
where $\eta_{i}\left(t\right)$ are independent Gaussian random variables
with mean zero and unit variance. The density matrix can be obtained
by averaging over the realizations of the noise,

\begin{align}
\hat{\rho}(t+dt)= & \overline{|\psi(t+dt)\rangle\langle\psi(t+dt)|},
\end{align}
where $|\psi(t+dt)\rangle=\hat{U}(t+dt,t)|\psi(t)\rangle$ and the
over-bar represents averaging over the noise realizations. In this
work, we set the noise strength to $\gamma=1$ and the time-step to,
$dt=0.1.$ We have corroborated that our results do not change if
$dt$ is further reduced. We average over $10$ noise trajectories
and over $10$ phase realizations ($\phi$ in Eq.~(\ref{eq:hamiltonian}),
see Ref.~\citep{lezama_mergold_love_logarithmic_2022} for further
numerical details). We have found this averaging sufficient to reduce
the statistical uncertainty of the data. To study the transport properties,
we consider the following observables:

\paragraph*{Particle transport at infinite temperature.}

It is easy to check that irrespective of $\hat{H}$ the RHS of Eq.~(\ref{eq:lindblad})
vanishes for Hermitian Lindblad jump operators and $\hat{\rho}\propto\mathbb{I}$,
such that any initial state will reach an infinite temperature state.
Therefore, to avoid transient effects we directly characterize particle
transport at infinite temperature. For this purpose we look at the
two-point density-density correlation function, 
\begin{equation}
C_{i}\left(t\right)=\text{Tr\,\ensuremath{\left[\hat{\rho}_{\infty}\left(\hat{n}_{i}\left(t\right)-\frac{1}{2}\right)\left(\hat{n}_{L/2}-\frac{1}{2}\right)\right]}},\label{eq:4}
\end{equation}
where $\hat{\rho}_{\infty}=\mathbb{I}/\mathcal{N}$ is the infinite-temperature
density matrix, and $\hat{n}_{L/2}$ corresponds to an initial density
excitation at the center of the lattice. For non-interacting systems,
Eq.~(\ref{eq:4}) can be written as
\begin{align*}
C_{i}\left(t\right)= & \left|\text{Tr}\left[\hat{\rho}_{\infty}\hat{a}_{i}^{\dagger}\left(t\right)\hat{a}_{L/2}\right]\right|^{2}=\frac{1}{4}\left|U_{i,L/2}^{s}\left(t,0\right)\right|^{2},
\end{align*}
where $U_{i,L/2}^{s}\left(t,0\right)$ is the single-particle propagator.
The second equality follows, since for unitary unraveling $\hat{a}_{i}^{\dagger}\left(t\right)$
can be written as: $\hat{a}_{i}^{\dagger}\left(t\right)=\sum_{k}U_{ik}^{s}\left(t,0\right)\hat{a}_{k}^{\dagger}$
and $\text{Tr}\left[\hat{\rho}_{\infty}\hat{a}_{i}^{\dagger}\hat{a}_{j}\right]=\frac{1}{2}\delta_{ij}$.
Since $U_{i,L/2}^{s}\left(t,0\right)$ has dimensions $L\times L$,
this allows us to efficiently study the system for large $L$.

We characterize the nature of transport using the root mean squared
displacement (RMSD) \citep{steinigeweg_2009,bar_lev_absence_2015,luitz_ergodic_2017,steinigeweg_real-time_2017},

\begin{align}
\tilde{R}\left(t\right)=\sqrt{\sum_{i=1}^{L}\left(i-\frac{L}{2}\right)^{2}C_{i}\left(t\right)} & .
\end{align}
Typically, the RMSD grows as a power law in time, $\tilde{R}\left(t\right)\sim t^{\alpha}$,
where the dynamical exponent $\alpha=1/2$ corresponds to diffusive
transport and $\alpha=1$ to systems with ballistic transport. Regimes
characterized by sub-diffusive and super-diffusive transport involve
exponents ranging between $0\leq\alpha<0.5$ and $0.5<\alpha<1$,
respectively. For localized systems $\alpha=0$.

\begin{figure}
\includegraphics[width=1\columnwidth]{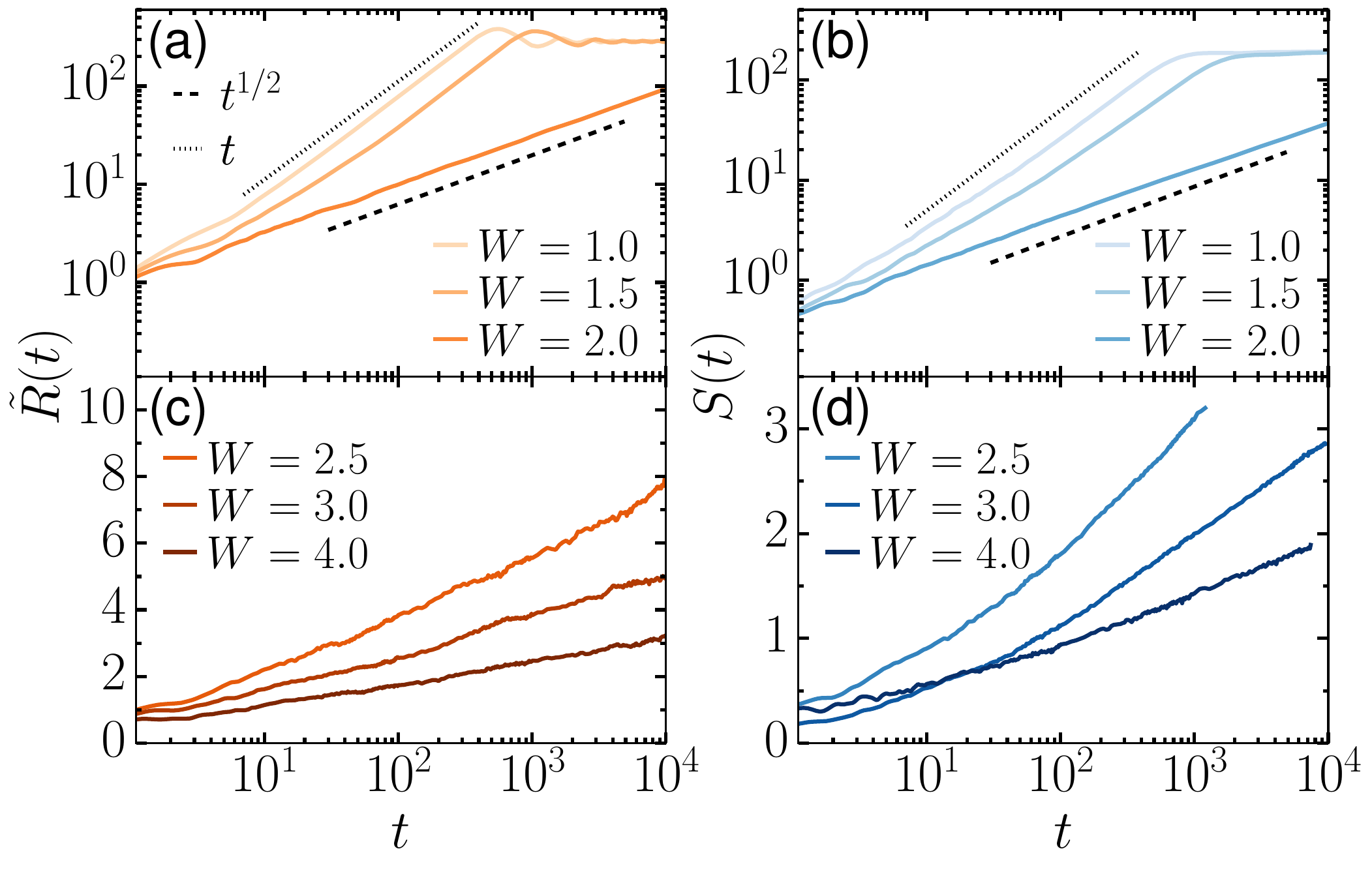}

\caption{Root mean-squared displacement $\tilde{R}\left(t\right)$ (left panels)
and entanglement entropy $S\left(t\right)$ (right panels) as a function
of time; for system size $L=1000$. The top panels, which are plotted
on a log-log scale, correspond to the delocalized phase and the critical
point. The bottom panels, which are plotted on a semi-log scale, correspond
to the localized phase. The black dashed and dotted lines provide
a guide to ballistic and diffusive transport, respectively. The statistical
error is of the order of the line width.}

\label{fig:fig2}
\end{figure}

\paragraph*{Dynamics of entanglement entropy.}

While the entanglement entropy is not a good measure of quantum information
for mixed states~\citep{Eisert_entanglement:rev}, for the unitary
unraveling, it is well defined for each of the trajectories. Therefore,
in situations when one can physically justify the specific form of
the unraveling \footnote{For example, a system in a time-dependent noisy potential},
the entanglement averaged over the various trajectories is a sensible
quantity \citep{Cao:2019,lezama_mergold_love_logarithmic_2022}. Another
advantage of the unitary unraveling is that the wavefunction $\left|\psi\left(t\right)\right\rangle $,
is Gaussian through the entire evolution, which allows an efficient
computation of the entanglement entropy, using the relation,

\begin{align}
S\left(t\right) & =-\sum_{\alpha}\left[c_{\alpha}\left(t\right)\text{ln }c_{\alpha}\left(t\right)+\left(1-c_{\alpha}\left(t\right)\right)\text{ln}\left(1-c_{\alpha}\left(t\right)\right)\right].
\end{align}
Here, $c_{\alpha}\left(t\right)$ are eigenvalues of the correlation
function $\langle\psi\left(t\right)\left|\hat{a}_{i}^{\dagger}\hat{a}_{j}\right|\psi\left(t\right)\rangle$,
restricted to the subsystem of interest~\citep{peschel_calculation_2003,peschel_reduced_2009}.
For ballistic transport, the entanglement entropy grows linearly with
time, while anomalous transport is characterized by a sub-linear growth~\citep{calabrese_2005,kim2013ballistic,luitz2016extended,lezama2019power}.

\section{Results\label{sec:Results}}

In this Section, we discuss our results for local coupling the heat
bath to one site of the system and to a finite fraction of system
sites. In finite fraction coupling we either couple the heat bath
to a central region of the chain or to equally separated lattice sites
through the entire chain (Fig.\ref{fig:sch}). All the results below
are obtained for a chain length of $L=1000.$

\subsection{\emph{Local coupling}}

In the left panels of Figure~\ref{fig:fig2} we plot the dynamics
of the root mean squared displacement for the delocalized, critical,
and localized phases of the AAH model. In the absence of coupling
to the heat bath, particle transport in these phases is ballistic,
anomalous and absent, respectively \citep{varma_fractality_2017,purkayastha_anomalous_2018}.
As can be seen from Fig.~\ref{fig:fig2}(a), local heat bath does
not affect the nature of transport in the delocalized and critical
regimes. On the other hand, it induces logarithmic transport in the
localized phase, similar to the case of the disordered Anderson model
\citep{lezama_mergold_love_logarithmic_2022}. Within the time range
considered, there is no sign of a crossover to diffusion, as opposed
to the case where the noise is coupled to \textit{all} the sites \citep{amir_classical_2009,znidaric_2010,znidaric_2013,znidaric_2017,gopalakrishnan_noise-induced_2017}.

In the right panels of Fig.~\ref{fig:fig2} we present the growth
of the entanglement entropy starting from a random product state.
We use a random product state so that initially the entanglement entropy
is zero, which provides us with a finite regime of entanglement entropy
growth. In the delocalized regime, the growth of the entanglement
entropy is unaffected by the presence of the local heat bath and is
consistent with a power-law dependence on time, observed in closed
systems \citep{roosz_nonequilibrium_2014}. In the localized case,
the heat bath induces logarithmic growth of entanglement entropy which
is reminiscent of the entanglement entropy growth in the MBL phase~\citep{znidaric2008many,bardarson2012unbounded,nanduri2014entanglement,znidaric2018entanglement}.
However, here this growth is accompanied with logarithmic particle
transport, while the presence of logarithmic particle transport in
the MBL phase is under debate~\citep{kiefer2020evidence,luitz2020absence,maximilian2022particle}.

\begin{figure}
\includegraphics[width=1\columnwidth]{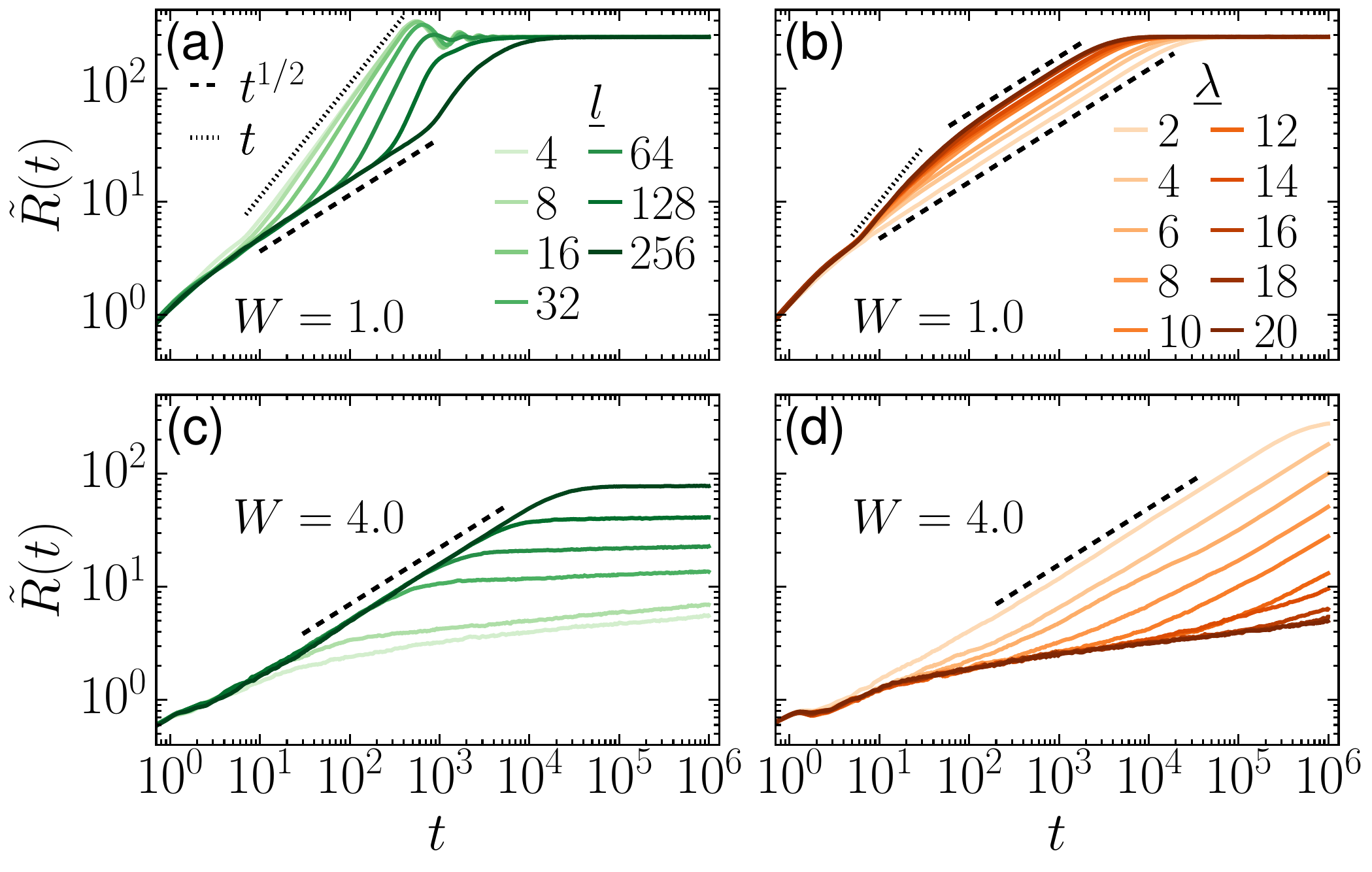}

\caption{Root mean-squared displacement $\tilde{R}\left(t\right)$, as a function
of time for $L=1000$. Left panels correspond to coupling of the heat
bath to a central region of length $l$. Right panels to coupling
the heat bath to system sites which are $\lambda$ distance apart.
Top panel correspond to $W=1$ and bottom panels to $W=4$. Dashed
and dotted lines are guides to the eye for diffusive and ballistic
transport, respectively. More intense colors stand for larger $\lambda$
or $l$.}

\label{fig:fig3}
\end{figure}

\begin{figure}
\includegraphics[width=1\columnwidth]{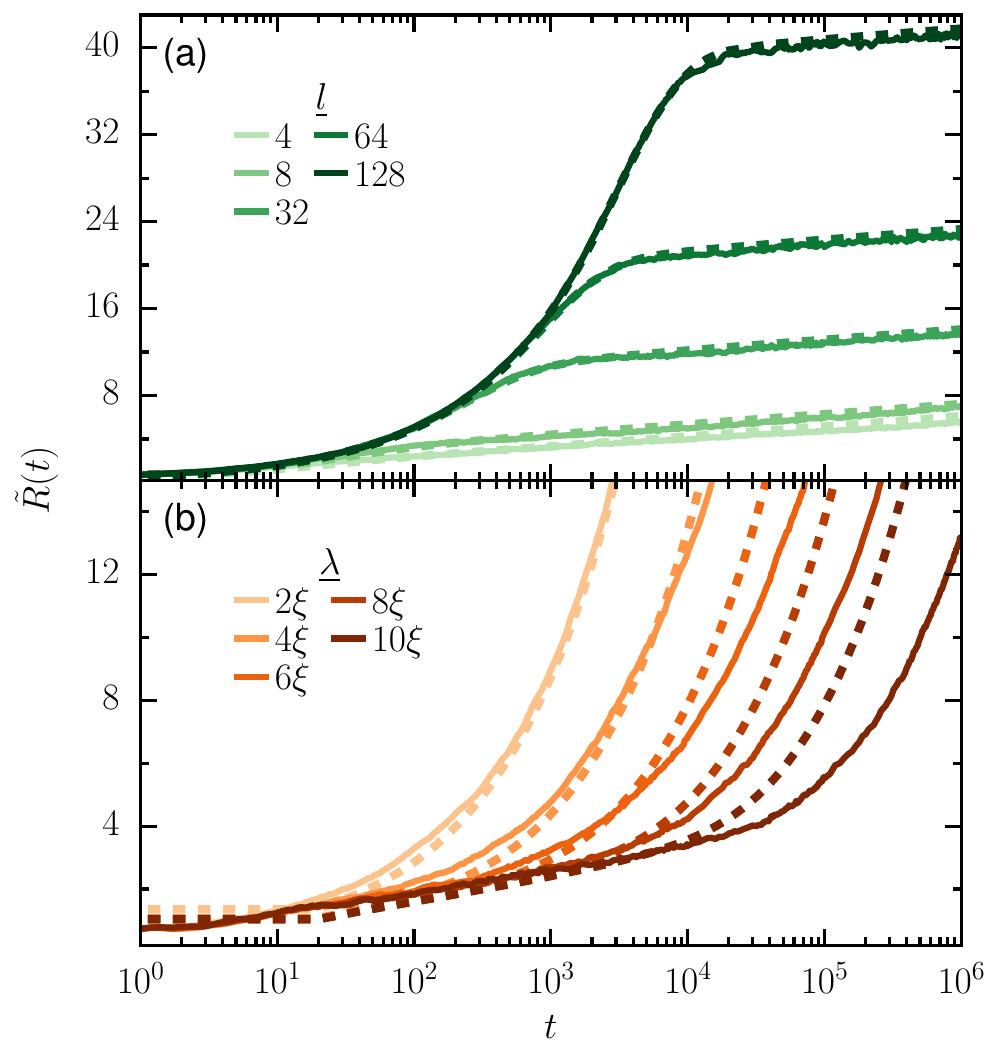}

\caption{Root mean-squared displacement $\tilde{R}(t)$, as a function of time;
for a number of coupled central region widths $l$ (top panel) or
the distance between the coupled sites, $\lambda$ (bottom panel).
Solid lines correspond to the the numerical solution of the Lindblad
equation (\ref{eq:lindblad}), and dashed lines to the solution of
the classical master equation (\ref{eq:phenomenological_master}),
with time rescaled by a factor of 20 to have the best fit with the
solid lines. The parameters used are $L=1000$ and $W=4$.}

\label{fig:fig4}
\end{figure}

\begin{figure}
\includegraphics[width=1\columnwidth]{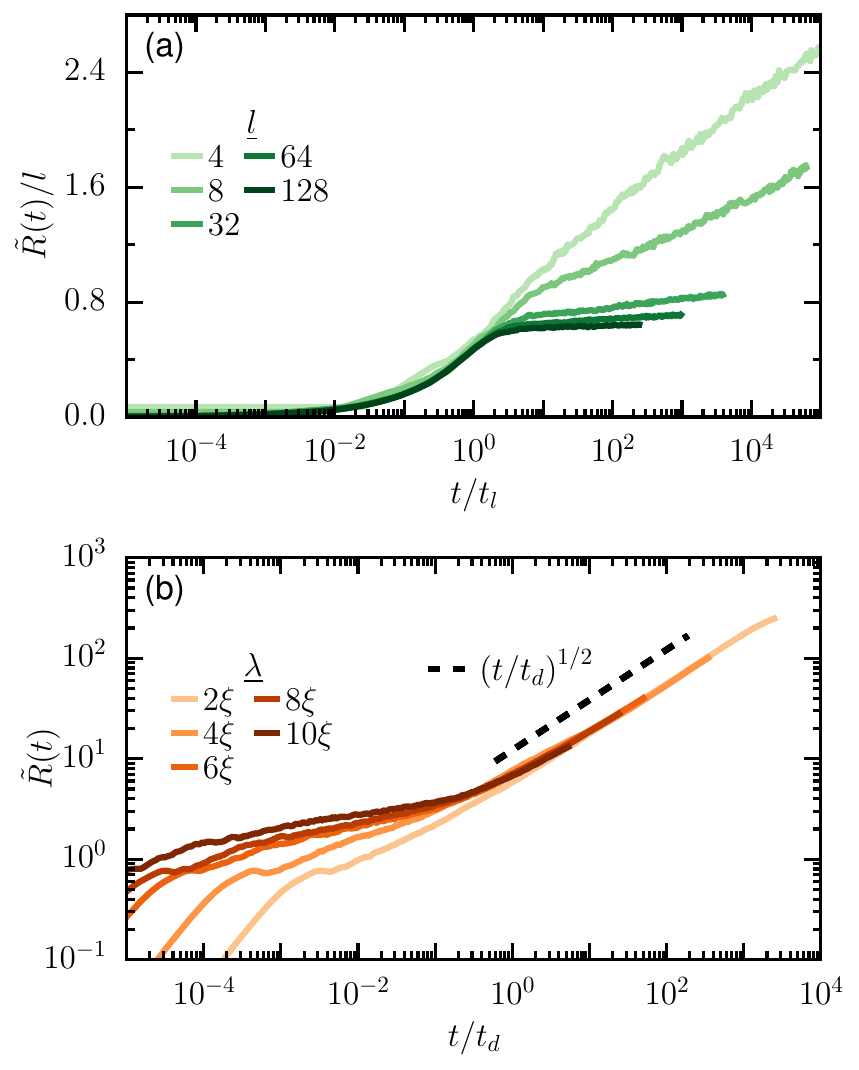}

\caption{Same as Fig.~\ref{fig:fig4} but with the time axis rescaled by the
crossover times $t_{l}$ or $t_{d}$ (see main text).}

\label{fig:fig5}
\end{figure}

\subsection{\emph{Coupling to a finite part of the chain}}

We have seen that local coupling to the heat bath is not affecting
transport in the delocalized phases and is inducing logarithmic transport
in the localized phase. In this section we study how transport is
affected when the heat bath is coupled to a finite fraction of system
sites. Moreover, we will show that the spatial configuration of the
coupled sites is important. Specifically we consider two different
configurations: coupling $l<L$ sites at the central region of the
chain, or coupling sites which are separated a distance $\lambda$
apart, such that their density is $1/\lambda$.

The left panels of Fig.~\ref{fig:fig3} show the dynamics of $\tilde{R}\left(t\right)$
in the delocalized phase (top panel) and the localized phase (bottom
panel) for different widths $l$ of the coupled central region. In
both delocalized and localized phases $\tilde{R}\left(t\right)$ initially
grows diffusively. After this initial diffusive growth it crosses
over to ballistic transport (Fig.~\ref{fig:fig3}(a)) in the delocalized
phase, or to logarithmic transport in the localized phase (Fig.~\ref{fig:fig3}(c)).
In both cases the crossover time, which we will designate by $t_{l}$,
grows with the width of the coupled region, $l$.

The right panels of Fig.~\ref{fig:fig3} show $\tilde{R}\left(t\right)$
for coupling sites which are a distance $\lambda$ apart from each
other. In the delocalized and localized phases, we observe a crossover
to diffusion. At the critical phase, transport remains practically
unaffected by the heat bath (not shown). In the delocalized phase,
the initial transport is ballistic (see Fig.~\ref{fig:fig3}(b))
and in the localized phase, the initial transport is logarithmic (see
Fig.~\ref{fig:fig3}(d)). In all cases the crossover time, which
we denote by $t_{d}$ increases with $\lambda$. In the next section
we introduce a classical master equation, which provides an explanation
to the dependence of the crossover times on the parameters of the
system.

\subsection{\emph{Classical picture}}

\begin{figure}
\includegraphics[width=1\columnwidth]{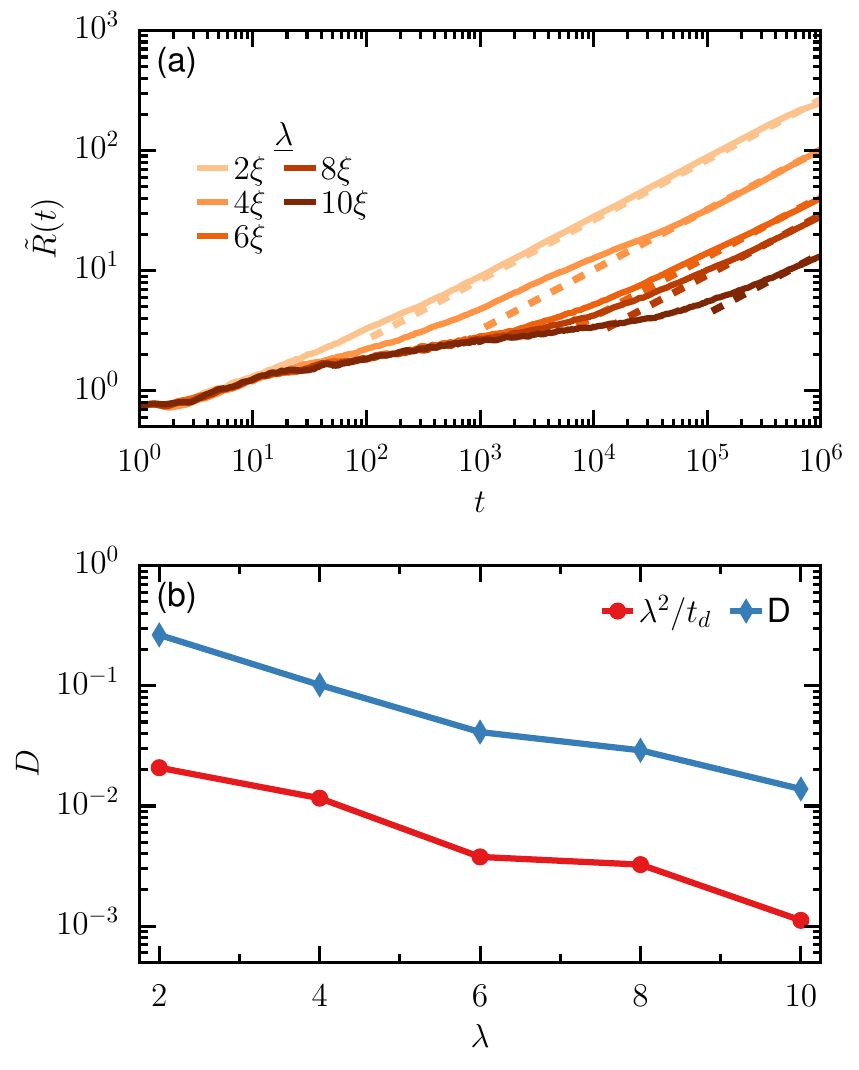}

\caption{Extraction of the diffusion coefficient for various $\lambda$ by
fitting $Dt^{1/2}$ to $\tilde{R}\left(t\right)$ obtained from the
solution of the Lindblad equation (dashed lines, top panel). The bottom
panel shows the diffusion coefficient as a function of $\lambda$
for $W=4$ and $L=1000$. The dashed blue line corresponds to the
theoretical prediction, with no fitting parameters.}

\label{fig:diffusion}
\end{figure}
The dephasing mechanism of the heat bath, diminishes the importance
of interference effects, and gives hope that classical treatment might
be sufficient to understand the underlying phenomenology. We therefore
follow the variable-range hopping approach of Mott \citep{mott160conduction},
and assume that coupling the system to the heat bath induces transitions
between localized single-particle eigenstates. The probability to
find a particle in a single-particle state $\alpha$, which we denote
by $p_{\alpha}$, evolves according to a classical master equation
\citep{amir_classical_2009,amir_localization_2010,fischer2016dynamics},

\begin{align}
\partial_{t}p_{\alpha}= & \sum_{\beta}\left(\Gamma_{\alpha\beta}p_{\beta}-\Gamma_{\beta\alpha}p_{\alpha}\right).\label{eq:phenomenological_master}
\end{align}
where the transition rates $\Gamma_{\alpha\beta}$ between the eigenstates
$\alpha$ and $\beta$ can be calculated from first-order perturbation
theory (see Appendix~\ref{sec:transition-rates}),

\begin{align}
\Gamma_{\alpha\beta}=\Gamma_{\beta\alpha} & =\gamma\sum_{k\in\text{coupled sites}}\left|\phi_{\beta}^{*}\left(k\right)\phi_{\alpha}\left(k\right)\right|^{2},
\end{align}
where $\gamma$ is the strength of the coupling, $\phi_{\alpha}\left(i\right)$
are the single particles states in the position basis, and the sum
$k$ runs over the sites coupled to the heat bath. In the localized
phase, by definition, $\phi_{\alpha}\left(i\right)$ decays exponentially.
We order the single-particle eigenfunctions such that $\phi_{\alpha}\left(i\right)$
has it center of mass around a site $\alpha$, which allows us to
approximate the transition rates as,

\begin{align}
\Gamma_{\alpha\beta}= & \gamma\sum_{k\in\text{coupled sites}}e^{-\left|\alpha-k\right|/\xi}e^{-\left|\beta-k\right|/\xi},\label{eq:master_rates}
\end{align}
where $\xi$ is the localization, which for the AAH model is $\xi=1/\ln\frac{W}{2}$~\citep{roosz_nonequilibrium_2014}.
The coupled site is, therefore, initiating transitions, or scattering,
between localized states within its neighborhood. Transitions to far-lying
states are exponentially suppressed with the distance from the coupled
site. For local coupling, this leads to logarithmic transport, $\tilde{R}\left(t\right)\sim\xi\ln\left(Jt\right)$
\citep{lezama_mergold_love_logarithmic_2022}.

When the heat bath is coupled to a region of final width $l$, the
transitions rates $\Gamma_{\alpha\beta}$ are approximately constant
in this region, and therefore a particle initiated in the coupled
region is expected to diffuse. It takes the particle time, $t_{l}\sim l^{2}$,
to leave this region. After leaving the coupled region, transitions
rates exponentially decay with the distance from the region, and transport
in the system is expected to be equivalent to the situation of local
coupling. When the coupled sites are at equal distances $\lambda$
apart, on a time-scale of moving between two nearby coupled sites,
transport is logarithmic, but at larger time-scales the expected motion
is diffusive. The crossover time can be obtained from $\lambda=\xi\ln\left(Jt_{d}\right)$,
yielding $t_{d}\sim J^{-1}\exp\left(\lambda/\xi\right)$, and the
diffusion constant as $D\sim\lambda^{2}/t_{d}=\lambda^{2}\exp\left(-\lambda/\xi\right)$.

In Fig.~\ref{fig:fig4} we compare the numerical solution of the
classical master equation (\ref{eq:phenomenological_master}) to the
numerical solution of the Lindblad equation (\ref{eq:lindblad}).
Since the classical rates (\ref{eq:master_rates}) are obtained phenomenologically,
the overall prefactor of the rates cannot be determined from microscopic
considerations. We determine it by rescaling the time axis of the
classical master equation such that the correspondence with the solution
of the Lindblad equation is optimal (\ref{eq:lindblad}). Apart from
this trivial rescaling of the units of time, there are no fitting
parameters. Remarkably, the agreement between the classical master
equation and the Lindblad equation goes \emph{beyond the qualitative
level} for the central coupled region (top panel). For equal spacing
coupling, there is still qualitative agreement, but the \emph{quantitative}
agreement is reasonable only for small $\lambda$/$\xi$.

In Fig.~\ref{fig:fig5} we test the predictions of the classical
theory for the crossover times, by rescaling of the time axis by $t_{d}$
or $t_{l}$, respectively. We see that such a rescaling correctly
identifies the crossover time for both couplings to the heat bath
when either $l$ or $\lambda$ are varied. The prediction for the
diffusion constant is verified in Fig.~\ref{fig:diffusion}. The
agreement is not quantitative, however the exponential decrease of
the diffusion constant with $\lambda$ is nicely captured.

\section{Discussion\label{sec:Conclusions-and-Discussions}}

Using the unitary unraveling of the Lindblad master equation, we study
the dynamical properties of the AAH model coupled to a dephasing heat
bath. We consider local coupling and coupling to a finite part of
the chain. This setup is partly motivated by dynamics in MBL systems
in the presence of finite density of ergodic bubbles \citep{luitz_bath:2017,deroeck2017stability,agarwal_rare-region_2017,khemani_critical:2017},
with the crucial difference that it is dissipative.

For local coupling of the heat bath in the delocalized and critical
phases of the AAH model, we didn't observe any qualitative effects
of the heat bath on the dynamics of the particle. On the other hand,
in the localized phase, the root mean square displacement, entanglement
entropy, and average energy (not shown) show asymptotic logarithmic
growth, as it occurs for the one-dimensional Anderson insulator in
the presence of a local noise \citep{lezama_mergold_love_logarithmic_2022}.
Suppose the region of the coupling to the bath is of finite width.
In that case, we find a regime of transient diffusion, which crosses
over to ballistic transport in the delocalized phase and logarithmic
transport in the localized phase. We have shown that this crossover
time increases as the square of the width of the region, as typical
for diffusion. When the heat bath is coupled to the system on equally
spaced sites, initial transport in the system is similar to the transport
with local coupling. Eventually, it crosses over to diffusion in all
phases. Specifically, in the localized phase, the crossover time increases
exponentially with the distance between the coupled sites.

We have shown that a classical master equation with transition rates
that exponentially decay from the location of the coupled sites captures
all the observed phenomenology. Moreover, it provides accurate predictions
of crossover times for all studied couplings of the heat bath. This
time scale is set by the time it takes for a particle to traverse
the distance $\lambda$ between two nearby coupled sites. In the delocalized
phase, this time is proportional to $\lambda$ (if the transport is
ballistic) or $\lambda^{2}$ (if it is diffusive). On the other hand,
it scales as $t_{d}\propto\exp\left(\lambda/\xi\right)$ in the localized
phase. Within the classical model, the motion of the particle between
the coupled sites can be viewed as a random walk with a spatial step
of $\lambda$ and a mean-free time of $t_{d}$, which means that the
motion is diffusive, with diffusion constant given by $D\sim\lambda^{2}/t_{d}=\lambda^{2}\exp\left(-\lambda/\xi\right)$.

In this work, we have focused only on fixed $\lambda$, such that
$D$ is also fixed. If $\lambda$ is allowed to vary randomly, such
that its distribution is unbounded, then the average time to transition
between coupled sites, $t_{d}$, can diverge. In this case, the average
diffusion coefficient will vanish, and transport will be subdiffusive
(see Refs.~\citep{taylor_subdiffusion_2021,turkeshi_destruction_2022}
and Appendix~\ref{sec:coupling-at-random-sites}). Since sites coupled
to a heat bath can be thought as perfect ergodic bubbles, we argue
that our results put an upper bound on transport in MBL systems due
to the presence of ergodic bubbles \citep{luitz_bath:2017,deroeck2017stability,agarwal_rare-region_2017,khemani_critical:2017}.

Here, we focus on a quasi-periodic model without mobility edges. In
the presence of mobility edges, at least in principle, the coupling
to the heat bath can create transitions between the localized and
delocalized states. Several interesting questions arise for these
models: would the coupling eliminate the intermittent logarithmic
transport regime? How will the dynamics depend on the initial conditions?
We leave these questions to future studies.
\begin{acknowledgments}
This research was supported by a grant from the United States-Israel
Binational Foundation (BSF, Grant No. $2019644$), Jerusalem, Israel,
and the United States National Science Foundation (NSF, Grant No.
DMR$-1936006$), and by the Israel Science Foundation (grants No.
527/19 and 218/19). D.S.B acknowledges funding from the Kreitman fellowship.
\end{acknowledgments}

\bibliography{AAH_noise}

\cleardoublepage{}

\selectlanguage{english}%

\appendix
\onecolumngrid
\selectlanguage{american}%

\section{Transition rates for white noise\label{sec:transition-rates}}

In this section, we calculate the transition rates using the first
order perturbation theory. Withing the unitary unraveling of the Lindblad
equation the dynamics of the system is described by a the time-dependent
Hamiltonian,

\begin{align*}
\hat{H}\left(t\right)= & \hat{H}_{0}+\hat{V}\left(t\right),
\end{align*}
where, $\hat{V}\left(t\right)={\displaystyle \sqrt{\gamma}\sum_{i}\eta_{i}\left(t\right)|i\rangle\langle i|}$
with $\eta_{i}\left(t\right)$ being a Gaussian random variable with
zero mean and unit variance. The noise is characterized by the correlation
function, $\langle\eta_{i}\left(t\right)\eta_{j}\left(t^{\prime}\right)\rangle=\delta_{ij}\delta\left(t-t^{\prime}\right)$.
The time evolution of the state $|\psi(t)\rangle$ in the eigenbasis
$|\alpha\rangle$ of $\hat{H}_{0}$ can be written as,
\begin{equation}
\partial_{t}c_{\alpha}=-i{\displaystyle \sum_{\gamma}\left\langle \alpha\left|\hat{V}\left(t\right)\right|\gamma\right\rangle c_{\gamma}\left(t\right)e^{-i\left(\epsilon_{\gamma}-\epsilon_{\alpha}\right)t}},
\end{equation}
where $\epsilon_{\alpha}$ is the eigenvalue of $\hat{H}_{0}$ which
corresponds to $|\alpha\rangle$. In the integral form, this corresponds
to,
\begin{equation}
c_{\alpha}\left(t\right)=c_{\alpha}\left(0\right)-i\int_{0}^{t}\sum_{\gamma}dt'\left\langle \alpha\left|\hat{V}\left(t\right)\right|\gamma\right\rangle c_{\gamma}\left(t\right)e^{-i\left(\epsilon_{\gamma}-\epsilon_{\alpha}\right)t'}.
\end{equation}
Back substitution of the LHS into the RHS and ignoring terms of order,
$O\left(V^{2}\right)$ gives,
\begin{equation}
c_{\alpha}\left(t\right)=c_{\alpha}\left(0\right)-i\int_{0}^{t}\sum_{\gamma}dt'\left\langle \alpha\left|\hat{V}\left(t\right)\right|\gamma\right\rangle c_{\gamma}\left(0\right)e^{-i\left(\epsilon_{\gamma}-\epsilon_{\alpha}\right)t'}.
\end{equation}
We assume that the system is initialized in state $\beta,$ such that
$c_{\gamma}\left(0\right)=\delta_{\gamma\beta}$ , which yields,
\begin{equation}
c_{\alpha}\left(t\right)=\delta_{\alpha\beta}-i\int_{0}^{t}dt'\left\langle \alpha\left|\hat{V}\left(t\right)\right|\beta\right\rangle e^{-i\left(\epsilon_{\beta}-\epsilon_{\alpha}\right)t'}.
\end{equation}
 The transition probability rate between state $|\alpha\rangle$ and
$|\beta\rangle$ is given by 
\begin{equation}
\Gamma_{\alpha\beta}=\left|c_{\alpha}\left(t\right)\right|^{2}/t,=\frac{1}{t}\int_{0}^{t}dt'\int_{0}^{t}dt''\left\langle \alpha\left|\hat{V}\left(t'\right)\right|\beta\right\rangle \left\langle \beta\left|\hat{V}\left(t''\right)\right|\alpha\right\rangle e^{-i\left(\epsilon_{\beta}-\epsilon_{\alpha}\right)\left(t'-t''\right)}.
\end{equation}
Now since averaging over the noise gives,
\begin{equation}
\overline{\left\langle \alpha\left|\hat{V}\left(t'\right)\right|\beta\right\rangle \left\langle \beta\left|\hat{V}\left(t''\right)\right|\alpha\right\rangle }=\gamma\sum_{i,j}\overline{\eta_{i}\left(t'\right)\eta_{j}\left(t''\right)}\left\langle \alpha|i\rangle\langle i|\beta\right\rangle \left\langle \alpha|j\rangle\langle j|\beta\right\rangle =\gamma\sum_{i}\left|\left\langle \alpha|i\right\rangle \right|^{2}\left|\left\langle \beta|i\right\rangle \right|^{2}\delta\left(t'-t''\right),
\end{equation}
we obtain

\begin{align*}
\Gamma_{\alpha\beta}= & \frac{\gamma}{t}\sum_{i}\left|\left\langle \alpha|i\right\rangle \right|^{2}\left|\left\langle \beta|i\right\rangle \right|^{2}\int_{0}^{t}dt'\int_{0}^{t}dt''\delta\left(t'-t''\right)e^{-i\left(\epsilon_{\beta}-\epsilon_{\alpha}\right)\left(t'-t''\right)}\\
= & \gamma\sum_{i}\left|\left\langle \alpha|i\right\rangle \right|^{2}\left|\left\langle \beta|i\right\rangle \right|^{2},
\end{align*}
where $\langle\beta|i\rangle=\phi_{\beta}^{*}\left(i\right)$ and
$\langle k|\alpha\rangle=\phi_{\alpha}\left(k\right)$.

\section{Coupling at random sites\label{sec:coupling-at-random-sites}}

\begin{figure}
\includegraphics[width=0.6\textwidth]{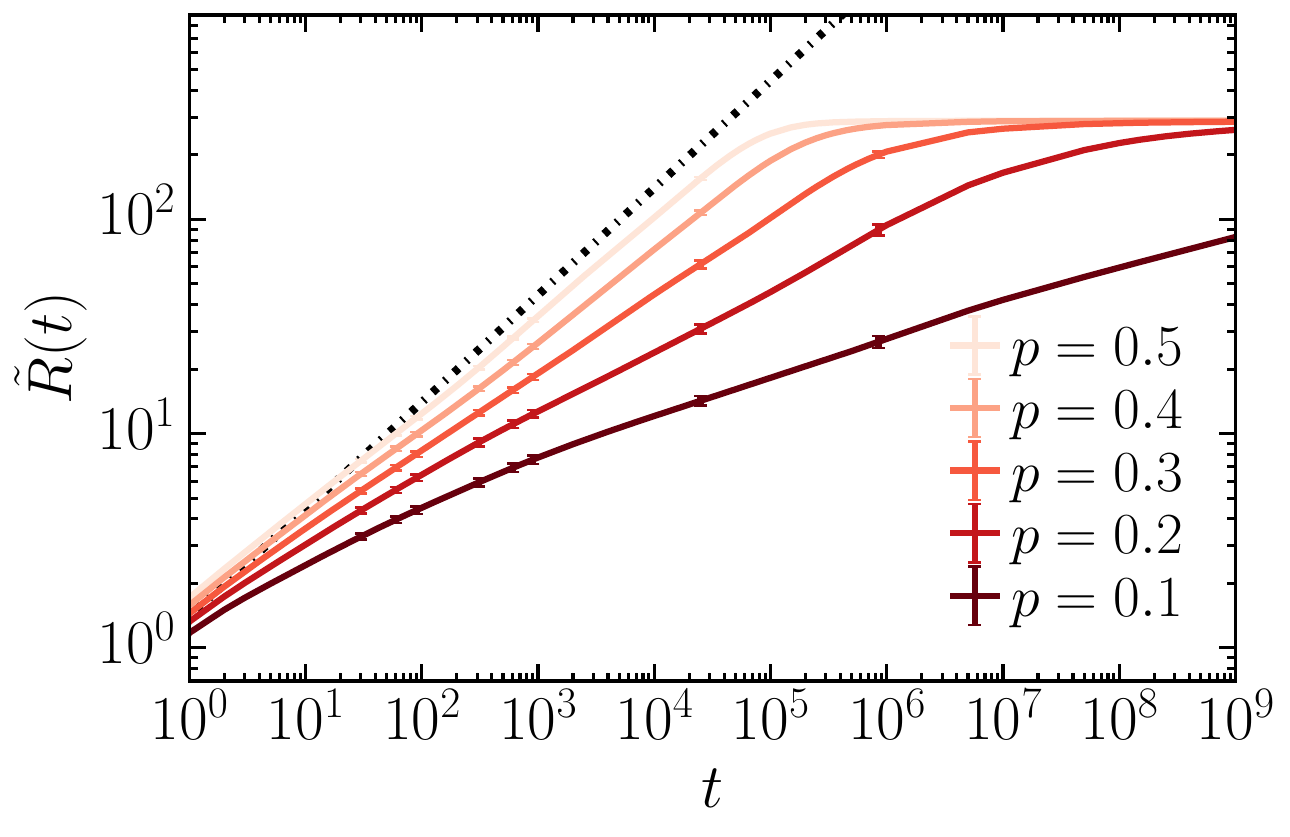}

\caption{Dynamics of $\tilde{R}\left(t\right)$ for for $W=4.0$, and the case
where each site of the system can be coupled with probability $p$.}

\label{fig:fig6}
\end{figure}

In this section, we consider the case where a given site is coupled
with probability $p$, which means that the distance between two coupled
sites is distributed according to the Poisson distribution. In Fig.~\ref{fig:fig6}
we shows the dynamics of $\tilde{R}\left(t\right)$ calculated using
the classical master equation for $W=4.0$ and and number of $p$.
We average the data over $50$ realizations of the couplings to the
heat bath. As opposed to the case where the coupled sites are placed
at an equal distance from each other leading to eventual diffusion,
here we find sub-diffusive transport with a dynamical exponent which
depends on $p$. Our results are consistent with Refs.~\citep{taylor_subdiffusion_2021,turkeshi_destruction_2022}
where the stationary state current is calculated in a similar system.

\end{document}